\documentstyle{aipproc}
\input epsf       

\begin{document}
\title{What can BeppoSAX tell us about short GRBs: An update from the Subsecond GRB Project}

\author{G.~Gandolfi$^{(1)}$ M.J.S.~Smith$^{(2),(4)}$ A.~Coletta$^{(2)}$ G.~Celidonio$^{(2)}$ L.~Di Ciolo$^{(2)}$ A.~Paolino$^{(2)}$ G.~Tarei$^{(2)}$ G.~Tassone$^{(2)}$ J.M.~Muller$^{(2),(4)}$ E.~Costa$^{(1)}$ M.~Feroci$^{(1)}$ F.~Frontera$^{(3),(5)}$ L.~Piro$^{(1)}$}
\address{$^{(1)}$IAS/CNR, Via del Fosso del Cavaliere, Roma, Italy\\
$^{(2)}$BeppoSAX-SOC/TELESPAZIO, Via Corcolle 19, Roma, Italy\\
$^{(3)}$TESRE/CNR, Via Gobetti 101, Bologna, Italy\\
$^{(4)}$SRON, Sorbonnelaan 2, Utrecht, The Netherlands\\
$^{(5)}$Dipartimento di Fisica, Universit\'a di Ferrara, Via del Paradiso 12, Ferrara, Italy}

\maketitle

\begin{abstract}
We present some statistical considerations on the BeppoSAX hunt for subsecond GRBs at the Scientific Operation Center. Archive analysis of a BATSE/SAX sub-sample of bursts indicates that the GRB Monitor is sensitive to short ($\le$ 
2 sec) events, that are in fact $\approx$ 22$\%$ of the total. The non-detection of corresponding prompt X-ray counterparts to short 
bursts in the Wide Field Cameras, in about 3 years of operations, is discussed: with present data no implications on the X-to-$\gamma$-ray spectra of short vs long GRBs may be inferred. Finally, the status of searching procedures at SOC is reviewed.
\end{abstract}

\section*{Introduction}

The nature and the origin of short bursts, namely  the events with a 
duration $\le$ 2 s, that clearly represent a separate class from longer ones 
(Kouvelioutou, 1993),  is one of the most interesting and puzzling open
problems in GRB studies. The celestial distribution of these events is 
clearly isotropic and their gamma spectrum predominantly harder than that 
of longer ones, while the two classes have the same peak flux range. A lot 
of questions naturally arise when considering the issue and comparing the 
phenomenology with the successful paradigm of Fireball models: are the 
short bursts caused by a different emission mechanism? Do they originate 
in extragalactic sources as well? Is their LogN-LogS euclidean (Tavani, 
1998)?  Are there other duration subclasses (Cline, 1999)? Do they possess 
an afterglow signature as the longer ones? At present no prompt 
counterpart detection at any wavelength is available and the recent 
afterglow discoveries refer only to the long bursts class, leaving a 
conspicuous gap in our knowledge of the phenomenon. Hence, the 
identification and localization of at least some event of this kind, that we 
call ``subsecond bursts'', remains a key objective of GRB research, an 
objective that can only be pursued at the moment by means of BeppoSAX 
satellite and, in the near future, by means of missions like Hete2, Integral 
and Swift.\newline %
\indent We present here some statistical consideration on BeppoSAX triggering 
sensitivity to short bursts with GRBM. Finally, the non detection of 
subsecond events in the sample of WFC selected GRBs is analysed and the 
new real-time procedure for on-ground triggering at Scientific Operation 
Centre is briefly resumed. 

\section*{Estimation of GRBM Efficiency in triggering Short Bursts}

The basic procedure for GRB counterpart detection at the BeppoSAX Scientific 
Operation Centre, performed since the very beginning of the mission, relies on the visual check of PDS Lateral Shield and WFC ratemeters around GRBM trigger times (Coletta, 1998). This procedure is being recently improved and extended, 
discriminating as possible the short, particle induced, GRBM fake events from real bursts (Feroci, 1999), monitoring the BATSE triggers as well and  performing an additional on-ground search for excesses of counts on Lateral shields and WFCs below the trigger threshold (Smith, 1999; Gandolfi, 1999). In this work we address the general problem of the BeppoSAX sensitivity to GRBs with duration less than 2 s and we try to test the significance of experimental results in the field, i.e. the non detection of subsecond bursts at quick look level. The first point to evaluate is the efficiency of GRBM in triggering short events, since the fundamental method used to identify GRBs - the only one systematically applied to data for a long time, is the check of the counts at trigger times both in WFC and LS lightcurves with 1 s bins. 
If the sample of BeppoSAX triggers doesn't include the majority of short bursts we would expect a priori a very low probability of finding subsecond X-ray counterparts, because the general SOC monitoring of WFC transients at a quick look level has a poor sensitivity to excesses of counts that last less than 8 s (the standard bin adopted to scan lightcurves on an orbital basis).
In order to estimate the GRBM sensitivity in the duration range of interest with respect to BATSE,  we have selected the subsample of common triggers in the period 10/11/1996 - 15/9/1998 (BeppoSAX/SOC Catalog, 1999; BATSE Current Catalog, 1999). 
This guarantees the reality of SAX events (and automatically discriminates ``fake'' triggers), that are also analyzed in high time resolution mode (8 ms bin) in order to  obtain the T90 (i.e. the time interval during which they emit the 90$\%$ of their fluence).  We find 111 common triggers, with 24 events under the 2 s threshold that corresponds to a 22$\%$ of the total. Comparing this value with the 26$\%$ of BATSE duration distribution and assuming that the probability is described by a binomial, we conclude that the number of short bursts detected is still compatible with completeness with a probability of 15$\%$, but is almost certainly affected by a bias for very brief events. 
In fact the bins with GRBs under 0.5 s exhibit a slight deficit of the order of 25$\%$ with respect to BATSE 4B duration distribution: coherently with expectations, the triggering efficiency is probably poor in that range due to the small fluence of shortest events (Fig. 1). An offline analysis of BATSE positions, WFCs field of view and ratemeters counts at the relevant trigger times confirms the non detection of counterpart in each one of the 24 cases.

\begin{figure}
\epsfxsize=9.0cm \centerline{\epsfbox{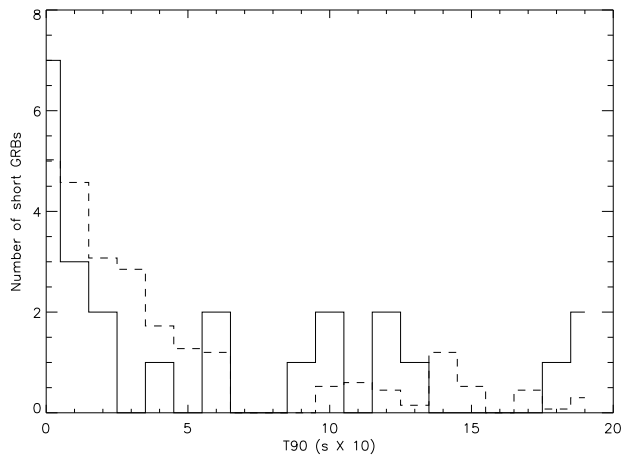}}
\caption{Duration distribution of short GRBs. The solid line is the common BeppoSAX/Batse Sample distribution, the dashed one is the normalized 4B Batse distribution, for comparison.}
\label{histogram}
\end{figure}

\section*{Increasing the Sensitivity with On-Ground Triggering: The BeppoSAX Subsecond Bursts Project}

The efficiency of the GRBM strongly depends from the spectral properties of the burst and from its maximum flux in the 1 s on-board trigger time bin. Other selection effects that could affect GRBs detection are mainly geometrical and depend on the effective area illuminated by the burst: in case of an event exactly on the axis of the LS (this is the most favorable condition to catch the counterpart in WFC field of view if involved lateral shield is unit 1 or unit 3) the area is minimum and the probability of triggering the GRBM is consequently minimized. For this reason an on-ground triggering procedure (part of the BeppoSAX subsecond GRB Project) has been implemented at SOC last summer (Gandolfi, 1999), with the aim of increasing the global detection efficiency of short and long-faint bursts. 
This semi-automatic procedure relies only on the comparison between lateral shields and WFC ratemeters in the whole orbit with a 1 s bin. The new trigger condition, that has been empirically chosen in order to achieve the best sensitivity to short events without overloading the amount of quick look work and is certainly more effective with respect to the on-board one, selects events which have a 3$\sigma$ excess of counts in the lateral shield and at the same time a 3.5$\sigma$ excess in the corresponding WFC ratemeter bin. The WFC threshold guarantees that all the bursts detectable in the celestial image, that is those with a global one bin fluence of at least 70 counts in the 2-10 keV range (Smith, 1999) are selected.
Interesting examples of short events detected by the new semi-automatic routine, are GRB991014 and GRB991106: the first is slightly too long to be considered a subsecond event, the latter is probably a X-ray/GRB without a detected signal in the GRBM. 
Other candidate short bursts have been found, but the WFC excess of counts never satisfied the above condition and the transients were in fact not detectable in the corresponding WFC celestial image.The detection of a number of such events, even if it doesn't allow to localize the burst, could help to constrain the spectral properties of short GRBs with statistical significance.

\section*{Experimental Results and Implications}

The present number of GRBs with X-ray prompt counterpart detected by BeppoSAX 
is 27 and no one is classified as short event. Is this result compatible with the expectations? If we assume the completeness in our GRBM  detections in the 
subsecond duration range (and we have seen that we are at least almost complete) we would expect to detect  the 26$\%$ of short  bursts also in the sample of GRBs with revealed counterpart, i.e. $\sim$7. This is not the case, but we must remember that just 17 of these events have been triggered by GRBM, the only reliable detection method if the X-ray counterpart has a very low fluence (the GRB quick look procedure guarantees an optimal analysis of WFC, at 1 s or less resolution, only around trigger times). 
Hence any statistical consideration should be based on the GRBM trigger catalog, which has been - and is being - carefully inspected in real time by Duty Scientists at SOC.
Furthermore, the most stringent limit to subsecond counterpart identification is the WFC ratemeter sensitivity in the 1 s bin inspected, that decreases with the decreasing duration of the transient. Considering the duration distribution of short GRBs and the X ray counterparts peak flux distribution for long bursts (Frontera, 1999; various IAUCs), we can estimate a rough global detection efficiency: with a minimum 3$\sigma$ signal in the 1 s bin of WFC ratemeter, in the hypotesis of an average offset of 10 degrees from the centre of the field of view and of a Crab-like spectrum for the source, we should miss about the 50$\%$ of expected events. 
The conversion from the peak flux to the fluence that is compared to the threshold in counts for the bin ($\sim$ 36 in a standard WFC empty field), is done assuming a constant and maximum emission during T90.
The expected number of subsecond counterparts in the GRBM-selected sample is 
therefore $N_{xs}$ = $f_{\gamma s}$ $\times$ $\eta$ $\times$ $N_x$ , where $f_{\gamma s}$ is the percentage of short gamma bursts, $\eta$ the estimated global detection efficiency for short X-ray counterparts and $N_x$ the global number of X ray counterparts detected by BeppoSAX GRBM. We find with all the above assumptions a value of  2 subsecond events. 
The non detection is not statistically significant, because considering again the binomial distribution (with p = $f_{\gamma s}$ $\times$ $\eta$ the probability of detecting the counterpart of a short GRB instead of a long one, x the GRBM-selected global number of counterpart detections, in this case no one, and n the dimension of the sample, that is 17 events) its probability P(x,n,p) corresponds to P(0,17,0.13) $\sim$ 0.09.
This result strongly encourages to increase the external trigger capability at quick look level in order to maximize the probability of finding a subsecond counterpart. 
In fact, extending to candidate events not selected by GRBM the real-time standard GRB procedure, with its robust and tested efficiency, will surely increase the chance of identifying low flux events.

\section*{Conclusions}

At present, no sure prompt X-ray counterparts of subsecond GRBs have been 
noticed in BeppoSAX WFCs celestial images and we have shown that 
GRBM triggered detections in this range are compatible with 
completeness, with a probable slight deficit at shortest durations. This 
implies, taking into account WFC sensitivity limits and assuming a X to $\gamma$ peak ratio similar to that of longer GRBs, that 2 events on a total of 17 
triggered with revealed counterpart, should be subsecond bursts. The non detection is clearly not sufficient to test on a statistical basis any spectral hypotesis (i.e. the correctness of the assumption on X to $\gamma$ peak ratio). 
Furthermore, we can sistematically rely on non triggered detections (that 
represents the 63$\%$ of the whole sample of discovered counterparts) just 
since about 6 months, thanks to the new on-ground triggering routine 
implemented at SOC. No inference can be made on the basis of the  entire 
counterparts catalog (i.e. $\sim$ 3 / 4 events of 27), because untriggered events with a prompt counterpart could have been easily missed at a quick look 
level with the old standard GRB procedure. On the other side, archive 
analysis of the GRBM triggers catalog and of the WFC ratemeters of the 
whole mission are in progress, and the optimization of quick look 
procedures to identify and analyze at best the high time resolution data has 
been now obtained within the frame of the BeppoSAX Subsecond Bursts 
Project.

\end{document}